\documentclass[conference]{IEEEtran}

\usepackage{amsmath,amsthm,amssymb}
\usepackage{graphicx}
\usepackage{stmaryrd}
\usepackage{blkarray}

\DeclareMathOperator{\F}{\mathbb F}
\DeclareMathOperator{\W}{\mathcal W}
\DeclareMathOperator{\wt}{wt}

\newcommand{\mF}{\mathcal F}
\newcommand{\set}[1]{\left\{{#1}\right\}}

\begin{document}

\title{Algebraic matching techniques for fast decoding of polar codes with Reed-Solomon kernel}
\author{Peter Trifonov}
\author{\IEEEauthorblockN{Peter Trifonov} 
\IEEEauthorblockA{Saint Petersburg Polytechnic University,
Russia\\Email: petert@dcn.icc.spbstu.ru}}

\maketitle

\begin{abstract}
    We propose to reduce the decoding complexity of polar codes with non-Arikan kernels by employing a (near) ML decoding algorithm for the codes generated by kernel rows. A generalization of the order statistics algorithm is presented for soft decoding of Reed-Solomon codes. Algebraic properties of the Reed-Solomon code are exploited to  increase the reprocessing order. The obtained algorithm is used as a building block to obtain a decoder for polar codes with Reed-Solomon kernel.
\end{abstract}

\section{Introduction}
Polar codes are a novel class of capacity-achieving error correcting codes \cite{arikan2009channel}. However, the  rate of polarization provided by the  Arikan kernel is quite low. This results in  poor finite-length performance of Arikan polar codes, unless improved decoding algorithms and code constructions are used \cite{tal2015list,trifonov2016polar}. The rate of polarization and scaling exponent of polar codes can be substantially improved by replacing the Arikan kernel with  larger matrices \cite{korada2010polar}.
In particular, the $q\times q$ Reed-Solomon (RS) kernel over $\F_q$ was shown to achieve the optimal scaling exponent and polarization rate \cite{mori2014source,pfister2016nearoptimal}.
However, decoding of polar codes with non-Arikan kernels remains a major challenge, except in the case of the binary erasure channel. 

Computing the probabilities of kernel input symbols can be implemented using extended trellises of the codes generated by kernel submatrices \cite{griesser2002aposteriori}. However, the complexity of this method is prohibitively high even for small binary kernels. An alternative approach is to exploit the relationship of the kernel with an appropriate Arikan matrix \cite{trifonov2014binary}. Although some complexity reduction is provided by this approach, its computational cost still remains extremely high.

 It was suggested in \cite{miloslavskaya2014sequentialBCH} to use a near-ML decoding algorithm for the
codes generated by the submatrices of a binary kernel to simplify
evaluation of the probabilities of kernel input symbols, and speed up sequential decoding of polar codes with binary kernels. In this paper we extend this approach to the case of RS kernels.

The proposed approach is based on the ideas of the box-and-match algorithm \cite{Valembois2004box}. However, we exploit the MDS property and algebraic structure of RS codes  to reduce the decoding complexity. The obtained algorithm is shown to outperform the Koetter-Vardy algebraic soft decision decoding algorithm and adaptive belief propagation method for RS codes of length 15, and enables significant decoding complexity reduction for the case of polar codes with $8\times 8$ RS kernel.

The paper is organized as follows. In Section \ref{sBG} RS and polar codes are reviewed, as well as the classical erasure decoding method for RS codes. A generalization of the order statistics  algorithm to the case of RS codes is presented in Section \ref{sRSDecoding}. Application of the proposed algorithm to the case of polar codes with RS kernels is discussed in Section \ref{sPolarDecoding}. Simulation results are provided in Section \ref{sNumRes}.
\section{Background}
\label{sBG}
\subsection{Reed-Solomon codes}
\label{sRSReview}
$(n=q-1,k,n-k+1)$ RS code over $\F_q$ is a cyclic code with generator polynomial $$g(x)=\prod_{i=0}^{n-k-1}(x-\alpha^{b+i}),$$
where $b$ is typically set to $1$. An extended $(q,k,q-k+1)$ RS code is a set of vectors $(c_{-1},c_0,\dots,c_{q-2}),$  where $c_{-1}=-\sum_{i=0}^{q-2}c_i$, and $(c_0,\dots,c_{q-2})$ is a codeword of $(q-1,k,q-k)$ RS code with $b=1$.

Consider transmission of codewords of $(q-1,k,d=q-k)$ RS code over the $q$-ary erasure channel. 
Let $ i_0,\dots,i_{t-1}$ be the indices of erased symbols, $0\leq t<q-k$. Their values can be recovered as \cite{CodePracticeEng}
\begin{equation}
\label{mForney}
y_{j}=\frac{X_j^{-b}\Gamma(X_j^{-1})}{\prod_{l\neq j}(1-X_lX_j^{-1})},
\end{equation}
where $X_j=\alpha^{i_j}$ are the erasure locators,
\begin{equation}
\label{mBCHKeyEquation}
\Gamma(x)=\Lambda(x)S(x) \bmod x^{d-1},
\end{equation}
$\Lambda(x)=\prod_{j=1}^t(1-X_jx)$ is the erasure locator polynomial, 
$S(x)=\sum_{i=0}^{d-2}S_ix^i$ is the syndrome polynomial, 
\begin{equation}
\label{mRSSyndrome}
S_i=\sum_{j=0}^{n-1}y_j\alpha^{(b+i)j}, 0\leq i<d-1,
\end{equation}
and $y_j$ are the received symbols, where  erasures are replaced with zeroes. With some modifications, this approach can be also used for decoding of extended RS codes \cite{CodePracticeEng}.
\subsection{Polar codes}
Let $F_l$ be a $l\times l$ non-singular matrix over $\F_q$. 
$(n,k)$ polar code is a set of vectors $c_0^{n-1}=u_0^{n-1}A_{m}$, where $A_m=B_{l,m}F_l^{\otimes m}$, $n=l^m$, $u_i=0, i\in \mF$, $\mF\subset\set{0,\dots,n-1}$ is a set of $n-k$ indices of frozen symbols,   and $B_{l,m}$ is the digit-reversal permutation matrix, which corresponds to the mapping $$\pi\left(\sum_{i=0}^{m-1}j_il^i\right)=\sum_{i=0}^{m-1}j_il^{m-1-i}, 0\leq j_i<l.$$

Matrix $A_m$ gives rise to synthetic symbol subchannels
\begin{align}
\label{mSubchannelDef}
W_m^{(i)}(y_0^{n-1},u_0^{i-1}|u_i)=\quad\quad\quad\quad\quad\quad\quad\quad\quad\quad\quad\quad\quad\nonumber\\
\frac{1}{q^{n-1}}\sum_{u_{i+1}^{n-1}\in \F_q^{n-i-1}}\prod_{j=0}^{n-1}W_0^{(0)}(y_j|(u_0^{n-1}A_m)_j),
\end{align}
where $W_0^{(0)}(y|c)=W(y|c)$ is the transition probability function of the underlying memoryless symmetric channel.
If $F_l$ is  not permutation-equivalent to an upper-triangular matrix, and the algebraic closure of its elements is equal to $\F_q$, then the capacities of subchannels $W_m^{(i)}$ converge with $m$ to 0 or 1 symbols per channel use \cite{mori2014source}.

The (extended) RS kernel is given by 
\begin{equation}
\label{mRSKernel}
F_l=\begin{pmatrix}
\alpha_{-1}^{l-1}&\alpha_{0}^{l-1}&\dots&\alpha_{l-2}^{l-1}\\
\alpha_{-1}^{l-2}&\alpha_{0}^{l-2}&\dots&\alpha_{l-2}^{l-2}\\
\vdots&\vdots&\ddots&\vdots\\
\alpha_{-1}^{0}&\alpha_{0}^{0}&\dots&\alpha_{l-2}^{0}
\end{pmatrix},
\end{equation}
where $\alpha_{-1}=0,\alpha_i=\alpha^i, 0\leq i<l-1$, $\alpha$ is a primitive element of $\F_q$. It can be seen that the last $k$ rows of $F_l$  represent a generator matrix of an $(l,k)$ RS code. 

It is convenient to define probabilities
\[W_m^{(i)}\set{u_0^i|y_0^{l^m-1}}=\frac{W_m^{(i)}(y_0^{n-1},u_0^{i-1}|u_i)P\set{u_i}}{W(y_0^{n-1})},\]
which can be recursively computed as
\begin{align}
\label{mGeneralPolarRecursion}
W_{\lambda}^{(lj+i)}\set{u_0^{lj+i}|y_0^{N-1}}=\quad\quad\quad\quad\quad\quad\quad\quad\quad\quad\quad\quad\quad\quad\quad\nonumber\\\sum_{u_{lj+i+1}^{lj+l-1}\in \F_q^{l-i-1}}\prod_{s=0}^{l-1} W_{\lambda-1}^{(j)}\set{(u_{lt}^{lt+l-1}F_l)_s,0\leq t\leq j|y_{\frac{N}{l}s}^{\frac{N}{l}s+\frac{N}{l}-1}},
\end{align}
where $N=l^\lambda$.
The successive cancellation decoding algorithm makes decisions 
\begin{equation}
\label{mSCDecisionRule}
\widehat{u}_i=\begin{cases} \arg \max_{u_i\in\F_q} W_{m}^{(i)}\set{\widehat u_0^{i-1},u_i|y_0^{n-1}},& i\not \in \mathcal F \\0,&\text{otherwise}.\end{cases}
\end{equation}
This algorithm requires $O(n\log n)$ evaluations of $W_{\lambda}^{(lj+i)}$.
However, computing these probabilities according to \eqref{mGeneralPolarRecursion} requires $O(q^ll)$ operations. This makes the actual decoding complexity of polar codes with non-Arikan kernels prohibitively high, except for very small $q$ and $l$.

\section{Decoding of RS codes}
\label{sRSDecoding}
Order statistics decoding (OSD)\ is an efficient soft decoding method for arbitrary binary linear block codes \cite{fossorier1995softdecision}. It identifies the most reliable information set (MRIS) corresponding to the received noisy sequence, and constructs a list of  codewords, which differ from  the hard decision vector in at most $t$ symbols on the MRIS, where $t$ is a parameter of the algorithm called reprocessing order. The OSD\ algorithm relies on the assumption that the hard decision vector has typically a small number of errors on the MRIS. Its complexity can be further reduced by considering codewords which differ from the hard decision vector in at most $2t$ positions on the $k+s$ most reliable symbols, where $k$ is the dimension of the code, and $s\leq n-k$ is a parameter of the algorithm \cite{Valembois2004box}. The positions of $s$ most reliable received symbols outside of the MRIS are called control band. 
 Such codewords can be obtained by performing  OSD with reprocessing order $t$, saving the obtained codewords in boxes corresponding to the values of their control band, and matching with those having the control band at  sufficiently  small Hamming distance.  We propose to exploit the MDS property of RS codes in order to avoid storing of multiple codewords. Furthermore, the algebraic erasure decoding method discussed in Section \ref{sRSReview} can be used to avoid costly Gaussian elimination.
\subsection{Algebraic matching}
Consider decoding of an $(l,k,l-k+1)$ RS code $\mathcal C$ over $\F_q$. Let $y_0^{l-1}$ be the output of a $q$-ary input memoryless channel, and let $W\set{c|y_i}, c\in \F_q,0\leq i<l$, be the corresponding symbol probabilities. The objective of the decoder is to find $$\hat c_0^{l-1}=\arg\max_{c_0^{l-1}\in \mathcal C}\prod_{i=0}^{l-1}W\set{c_i|y_i}.$$ 
Let $x_i=\arg \max_{x\in\F_q}W\set{x|y_i} $ be the hard decision values corresponding to channel output. Let $L_{i}[z]=\log \frac{W\set{x_i|y_i}}{W\set{z|y_i}}, z\in \F_q$, be the log-likelihood ratios. Then the decoding problem can be equivalently stated as $$\widehat c_0^{l-1}=\arg\min_{c_0^{l-1}\in\mathcal C}EW(c),$$
where the ellipsoidal weight is defined as $$EW(c)=\sum_{i=0}^{l-1}L_i[c_i].$$
Let the reliability of a symbol be defined as $r_i=\min_{x\neq x_i}L_i[x]$, and let us assume that tuples $(L_i[x],i,x),$ where $x\neq x_i, 0\leq i<l$, are arranged in the ascending order of $L_i[x]$. Let $T_j$ be the $j$-th tuple in the ordered sequence, and let $T_j[s]$ be the $s$-th element of a tuple, $0\leq s<3$.

 Let us assume without loss of generality that $r_i\geq r_{i+1}, 0\leq i<l-1$. Let $c^{(0)}$ be a codeword, such that $c_i^{(0)}=x_i, 0\leq i<k$.

Order-$t$ reprocessing consists in enumerating all possible tuples $\mathbb T=(T_{j_1},\dots,T_{j_t}), j_1<j_2<\dots<j_t$, such that $T_{j_a}[1]\neq T_{j_b}[1]$ for all $a\neq b$, $0\leq T_{j_s}[1]<k$, and constructing codewords, such that
$$c_i=\begin{cases}
T_{j_s}[2],&\text{if $i=T_{j_s}[1]$ for some $1\leq s\leq t$}\\
x_i,&\text{otherwise}
\end{cases}$$
Let $P(\mathbb T)=(c_0,\dots,c_{k-1})-(x_0,\dots,x_{k-1})$
be the test pattern associated with tuple $\mathbb T$. 
Following \cite{Valembois2004box}, we define  $\sigma_P=(c_k,\dots,c_{k+s-1})$ as the $S$-complement of a codeword corresponding to test pattern $P$. It is possible to show that  for any test pattern $P'$, such that $P+P'$ differs in most $2t$ positions from $c^{(0)}$ over $k+s$ most reliable symbols, one has 
\begin{equation}
\label{mCBWeightBound}
\mathbf W(\sigma_{P+P'})\leq 2t-\mathbf W(P)-\mathbf W(P'),
\end{equation}
where $\mathbf W(z)$ denotes the Hamming distance of vector $z$ from the corresponding part of the hard decision vector $x_0^{l-1}$.

Instead of enumerating previously stored patterns $P'$ satisfying \eqref{mCBWeightBound}, we propose to explicitly construct them for each  pattern $P(\mathbb T)$. Namely, given a test pattern $P(\mathbb T)$ , we assume that the control band has at most $w=2t-\wt(P(\mathbb T))-\tau$ errors, and construct patterns $P(\mathbb T)+P':\mathbf W(P')\leq \tau$. In order to avoid construction of duplicate patterns, we assume that $\tau\leq \mathbf W(P(\mathbb T))$, and all non-zero elements of $P'$ are located in positions $i>T_{j_1}[1]$. To do this, we consider all tuples $(i_1,\dots,i_\tau): T_{j_1}[1]<i_1<\dots<i_\tau$, and all $\tau$-subsets $\set{h_1,\dots,h_\tau}\subset \set{k,\dots,k+s-1}$. For each combination of them we construct a codeword $\widetilde c$, such that  $$\widetilde c_i=\begin{cases}
c^{(P(\mathbb T))}_i,&i\in\set{0,k-1}\setminus{\set{i_1,\dots,i_\tau}}\\
x_i,&i\in\set{h_1,\dots,h_\tau},
\end{cases}$$
where $c^{(P(\mathbb T))}$ is the codeword corresponding to test pattern $P(\mathbb T)$. For each of the constructed codewords $c^{(P(\mathbb(T))+P'}$ we verify that it has indeed at most $w$ errors on the control band,  and keep the obtained codeword as a tentative decoder decision $\widetilde c$, if its ellipsoidal weight less than the ellipsoidal weight of the previous tentative decision. 

This can be done as long as $s\geq t$ using the erasure decoding algorithm discussed in Section \ref{sRSReview}. If $2\tau\leq s$, then it may be possible to use a classical algebraic error-correcting algorithm for $(k+s,k,s+1)$ RS code in order to find $P'$ without explicit enumeration of $i_1,\dots,i_\tau,h_1,\dots,h_\tau$.

It can be seen that the proposed algorithm 
considers at most ${k\choose t}(q-1)^t+{s\choose t}{k\choose t}$ test patterns. It can be implemented with memory of size $O((t+1)l)$, while the  Box-and-Match algorithm in the $q$-ary case  requires storage of size  $O(q^{s}+{k\choose t}q^t)$.

\subsection{Conditions of optimality}
\label{fCondOpt}
The conditions of optimality considered in \cite{Valembois2004box}
can be immediately re-used in the above presented algorithm.

Namely, for a test pattern $P$ of weight $j$ the obtained codeword 
may differ from the hard decision vector in at least  $\delta=\max(d-w-j,0)$ on positions outside of the MRIS, where $w=d_H(\widetilde c,x_0^{l-1})$, $d$ is the minimum distance of the code, and $\widetilde c$ is the tentative decoder decision. Hence, the necessary condition of optimality is 
$$\sum_{i:P_i\neq 0} L_i[P_i]+\sum_{j=l-\delta}^{l-1}r_j<EW(\widetilde c).$$  

Furthermore, if $(T_{j_1},\dots,T_{j_{t-1}},T_{j_t})$ 
fails the necessary condition of optimality, then all
 $(T_{j_1},\dots,T_{j_{t-1}},T_{j_t'})$ 
with $j_t'>j_t$ can be also excluded from the consideration.
More generally, if one has a global lower bound on the ellipsoidal weight of not-yet-tested patterns, then decoding can be terminated as soon as it exceeds $EW(\widetilde c)$.

\subsection{Fast re-encoding}
The proposed algorithm requires one to recover a codeword from the symbols in an arbitrary MRIS. Typically, in OSD-like algorithms this is implemented by performing Gaussian elimination on the generator matrix of the code, so that the identity submatrix is obtained in the columns corresponding to the MRIS. The complexity of this approach is $O(k^2n)$.  However,
one can use the techniques presented in Section \ref{sRSReview} to construct such a matrix. This involves the following steps:
\begin{enumerate}
\item Construct the locator polynomial $\Lambda(x)$ for symbols outside of the MRIS.
\item For $0\leq i<k$ do:
\begin{enumerate}
\item Let $p_i$ be the $i$-th most reliable symbol. Compute its syndrome polynomial as $S(x)=\sum_{j=0}^{d-2}\alpha^{p_i(j+b)}x^j$.
\item Let $G_{i,p_i}=1, G_{i,p_j}=0, j\neq i, 0\leq i,j<k$.
\item Compute $G_{i,p_j}$ from \eqref{mForney}, $k\leq j<n$.
\end{enumerate}
\end{enumerate}
The complexity of this algorithm is $O(k(l-k)^2)$.
\section{Decoding of polar codes}
\label{sPolarDecoding}
In order to reduce the decoding complexity of polar codes with non-Arikan kernels we propose to approximate the  sum in  \eqref{mSubchannelDef} with the maximal summand. The same transformation can be performed with $W_m^{(i)}\set{u_0^i|y_0^{l^m-1}}$.
That is, one obtains 
\begin{align}
W_m^{(i)}\set{u_0^i|y_0^{l^m-1}}\approx \W_m^{(i)}\set{u_0^i|y_0^{l^m-1}}=\quad \quad \quad \quad \quad \nonumber\\
\max_{u_{i+1}^{n-1}\in \F_q^{n-i-1}}\prod_{j=0}^{n-1}W_0^{(0)}(y_j|(u_0^{n-1}A_m)_j),
\end{align}
and the recursion in \eqref{mGeneralPolarRecursion} becomes
\begin{align}
\label{mGeneralPolarRecursionApprox}
\W_{\lambda}^{(lj+i)}\set{u_0^{lj+i}|y_0^{N-1}}=\quad\quad\quad\quad\quad\quad\quad\quad\quad\quad\quad\quad\quad\quad\quad\nonumber\\\max_{u_{lj+i+1}^{lj+l-1}\in \F_q^{l-i-1}}\prod_{s=0}^{l-1} \W_{\lambda-1}^{(j)}\set{(u_{lt}^{lt+l-1}F_l)_s,0\leq t\leq j|y_{\frac{N}{l}s}^{\frac{N}{l}s+\frac{N}{l}-1}},
\end{align}
where $\W_0^{(0)}\set{c|y}=W\set{c|y}$. The above expression can be recognized as computing the probability of a MAP codeword in the coset given by $u_{lj}^{lj+i}$ of the $(l,l-i)$ code generated by rows $i+1,\dots,l-1$ of matrix $F_l$. Hence, assuming the uniform distribution of $u_{lj+i+1}^{lj+l-1}$,  \eqref{mGeneralPolarRecursionApprox} can be computed using an arbitrary (near) ML\ decoding algorithm for that code. 
Observe that we do not need to construct  explicitly a  codeword of the corresponding coset. Only its probability $\W_{\lambda}^{(lj+i)}\set{u_0^{lj+i}|y_0^{N-1}}$ needs to be computed. 

In practice it is more convenient to implement decoding using the log-likelihood ratios $$E_\lambda^{(i)}(u_0^{(i-1)},u_i|y_0^{N-1})=\log \frac{\W_\lambda^{(i)}(u_0^{(i-1)},u_i|y_0^{N-1})}{\W_\lambda^{(i)}(u_0^{(i-1)},\widehat u_i|y_0^{N-1})},$$
where $\widehat u_i=\arg \max_{u_i} \W_\lambda^{(i)}(u_0^{(i-1)}, u_i|y_0^{N-1})$. 
This approach can be considered as a generalization of the min-sum approximation widely used for decoding of polar codes with Arikan kernel.
 
In the case of the RS kernel we propose to implement these calculations using the above presented algebraic matching algorithm. In this case one obtains  $$E_\lambda^{(i)}(u_0^{(i-1)},u_i|y_0^{N-1})=-EW(\widehat c),$$ where  $\widehat c$ is a codeword obtained by the decoder of the RS code. Observe that several simplifications are possible:
\begin{enumerate}
\item Sorting of  $T_j$ and $r_i$ can be performed only once for each kernel instance.
\item SC decoding of a polar code essentially reduces to decoding in a family of nested RS codes. At step $i$, instead of constructing the generator matrix from scratch, one can update the generator matrix obtained at step $i-1$. To do this, one needs to compute only $G_{s,p_{l-1-i}}, 0\leq s<l-i,$ and update the check part of the matrix as $G_{s,p_j}:=G_{s,p_j}+G_{s,p_{l-1-i}}G_{l-i,p_j}, l-i\leq j<l$. Hence, the cost of generator matrix update is only $(l-i)i$ operations.
\item Instead of employing the RS decoder for each $u_{lj+i}\in \F_q$ in \eqref{mGeneralPolarRecursionApprox}, one can apply the above described algorithm to a single vector $u_0^{lj+i-1}$, and for each codeword examined by it compute the corresponding value of $u_{lj+i}$ and the associated ellipsoidal weight. Then the LLR for each $u_{lj+i}\in \F_q$ is given by the smallest obtained ellipsoidal weight.
\end{enumerate}

\section{Numeric results}
\label{sNumRes}
\begin{figure}
\includegraphics[width=0.5\textwidth]{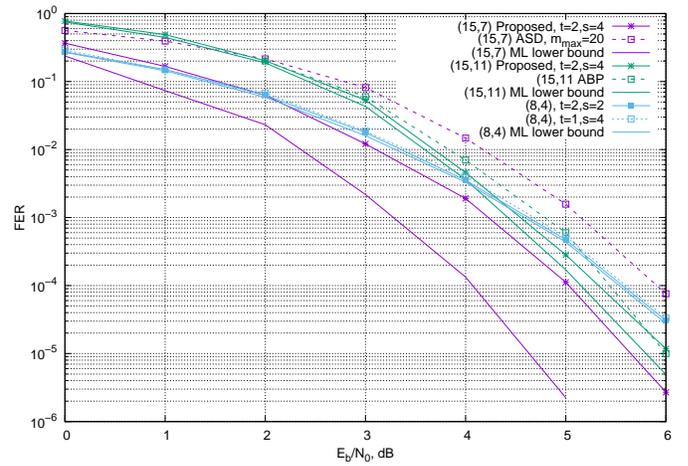}
\caption{Performance of short RS codes under the proposed decoding algorithm}
\label{fRSPerf}
\end{figure}

The performance of the proposed algorithms was investigated for the case of transmission of the binary image of RS and polar codes over AWGN channel. Figure \ref{fRSPerf} illustrates the performance of some short RS codes in the case of the proposed algorithm. For comparison, we report also the results for the Koetter-Vardy algebraic soft decision decoding algorithm \cite{koetter2003algebraic}, as well as for the non-binary adaptive belief propagation decoding, reproduced from \cite{bimberg2010performance}.
\begin{figure}
\includegraphics[width=0.5\textwidth]{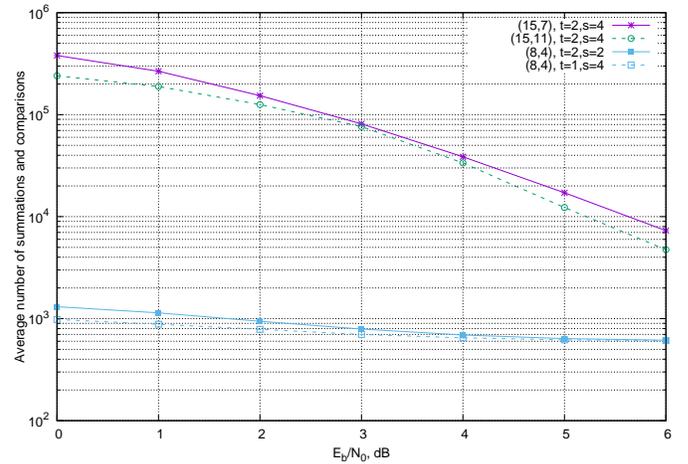}
\caption{Average complexity  of decoding RS codes}
\label{fRSComplexity}
\end{figure}
It can be seen that the proposed algorithm outperforms the ASD and ABP algorithms for codes of length 15, but still does not provide maximum likelihood decoding.
However, for the $(8,4,5)$ code  it provides near-ML performance even for reprocessing order $t=1$. Figure \ref{fRSComplexity} illustrates the average number of summation and comparison operations required by the proposed algorithm. It can be seen that it quickly decreases with SNR. 

\begin{figure}
\includegraphics[width=0.5\textwidth]{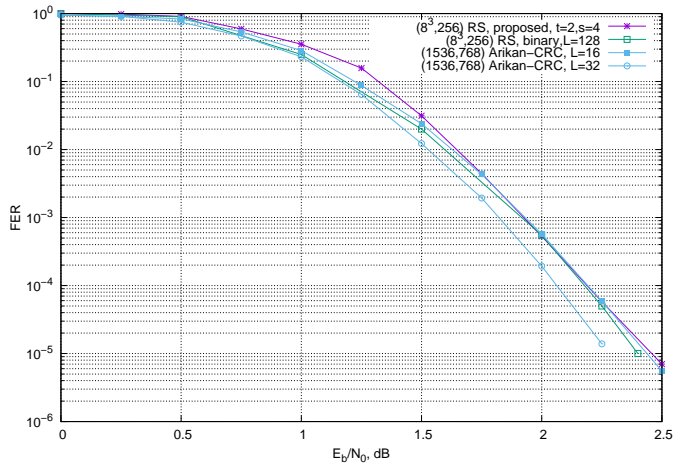}
\caption{Performance of polar codes with RS kernel}
\label{fPolar}
\end{figure}

Figure \ref{fPolar} illustrates the performance of a $(512,256)$ polar code with $8\times 8$ RS kernel constructed for $E_b/N_0=2$ dB. The symbol error probability for subchannels $W_m^{(i)}$  was obtained by simulations.
  For comparison, we report also the performance of the same code under the binary decoding algorithm introduced in \cite{trifonov2014binary}, which computes the exact values of probabilities $W_m^{(i)}$ given by \eqref{mGeneralPolarRecursion}. The latter algorithm was implemented in probability domain using the FFT\ for fast computation of the multi-dimensional convolutions.
 It can be seen that the proposed decoding algorithm provides 
near-optimal performance.   For comparison, we report also  the performance of a shortened $(1536,768)$ polar code with Arikan kernel and CRC-16 under list SC decoding. It can be seen that the polar code with RS kernel provides approximately the same performance  
as a polar code with CRC under list-16 decoding

\begin{figure}
\includegraphics[width=0.5\textwidth]{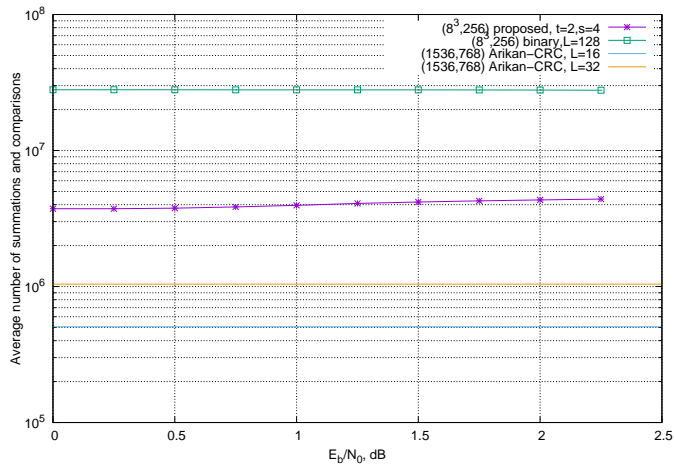}
\caption{Decoding complexity of polar codes with RS kernel}
\label{fPolarComplexity}
\end{figure}

Figure \ref{fPolarComplexity} presents the average complexity of the  decoding algorithms for polar codes with RS kernel. The proposed algorithm makes use of summations and comparisons, while the binary algorithm employs summations and multiplications. It can be seen that the proposed approach requires 7.5 times less operations than the binary algorithm. However, the complexity of the proposed algorithm increases slightly with SNR. The reason for this is that in \eqref{mGeneralPolarRecursionApprox} one needs to find the most probable vector $u_{lj+i+1}^{lj+l-1}$ not only for the most probable prefix $u_{lj}^{lj+i}$, but for all other values $u_{lj+i}\in \F_q$. This significantly reduces the efficiency of conditions of optimality discussed in Section \ref{fCondOpt}.

It can be also seen that the complexity of SC decoding of the polar code with $8\times 8$ RS kernel is only 8 times higher than the complexity of list-16 decoding of a polar code with Arikan kernel.
\section{Conclusions}
In this paper a novel decoding algorithm was presented for polar codes with RS kernel. The proposed approach relies on a near-maximum likelihood decoding algorithm for the codes generated by kernel submatrices. A generalization of the order statistics algorithm for $q$-ary codes was presented. The MDS property of RS codes was used in order to significantly reduce memory requirements of the obtained algorithm, and the algebraic structure of RS codes was exploited in order to reduce the complexity of the generator matrix transformations. 

The proposed algorithm was shown to provide near-optimal performance for polar codes with $8\times 8$ RS kernel and RS codes of length $8$. For the case of polar codes with RS kernel, the proposed approach was shown to have lower average decoding complexity compared to the binary decoding method. 

 For longer RS codes obtaining near-ML performance would require employing reprocessing order $t>2$, although the proposed algorithm outperforms algebraic soft-decision and adaptive belief propagation methods even for $t=2$. However, implementation of the successive cancellation decoder for polar codes with RS kernel requires near-ML\ decoding algorithms for the codes generated by kernel submatrices. Any progress in the development of such algorithms for RS codes would pave the way for efficient decoding of polar codes with larger RS kernels. 

\bibliographystyle{ieeetran}

\end{document}